# Spin-Polarized Standing Waves in the Fermi Surface of a Ferromagnetic Thin Film

J. Schäfer[1], M. Hoinkis[1], Eli Rotenberg[2], P. Blaha[3], and R. Claessen[1]

[1]*Institut für Physik, Universität Augsburg, 86135 Augsburg, Germany*
[2]*Advanced Light Source, Lawrence Berkeley National Laboratory, Berkeley, CA 94720, USA*
[3]*Institut für Materialchemie, Technische Universität Wien, A-1060 Wien, Austria*
(Received Day Month 2005)

The spin-selective electron reflection at a ferromagnetic-paramagnetic interface is investigated using Fe films on a W(110) substrate. Angle-resolved photoemission of the majority and minority Fermi surfaces of the Fe film is used to probe standing wave formation. Intense quantum well states resulting from interfacial reflection are observed exclusively for majority states. Such high spin polarization is explained by the Fermi surface topology of the connecting substrate, and we argue that Fe/W is a particularly suitable interface for that purpose.

In the pursuit of designing magnetoelectronics, the properties of magnetic solids are exploited such that the conductivity for a certain spin orientation is enhanced over its counterpart. Efficient spin valves are of high technological interest for data processing and information storage devices. One example is the giant magnetoresistance (GMR) effect established in Fe/Cr/Fe layers [1,2] as well as in Co/Cu/Co layers[3] where the second ferromagnetic layer serves as spin analyzer layer that affects the transmission properties of the device. More fundamentally, it is desirable to understand the propagation of spin-oriented electron wave functions across an interface as well as through bulk solids and how it can be controlled.

Spin-dependent transmission from a ferromagnet into a neighboring layer can be affected by scattering due to i) structural imperfections, ii) bulk interactions, and iii) the interfacial potential step for given spin. It remains a matter of debate which process is beneficial for a large GMR effect [4,5,6]. Structural influences [7] can be minimized in the film growth, yet defects may even be desirable. Second, spin-dependent bulk scattering relies on available empty states for electron-hole pair formation [5] and can lead to a spin-asymmetry in the mean free path. Third, the role of the potential step at an interface has been recognized theoretically [4,6], while experimental work has focused on magnetic coupling between two ferromagnetic layers [8].

Surprisingly, spectroscopic evidence on the spin-selectivity of interfacial reflection is scarce. Available data are limited to the nonmagnetic side of the interface. Extension of electron states from bulk Co into a Cu film has been reported from spin-polarized photoemission [9], yet with a rather limited degree of spin polarization. An experiment documenting spin-dependent reflection of conduction electrons in the ferromagnetic film itself is still lacking.

The continuation of Fermi level wave functions on both sides of the interface, as is relevant for transport, is indicated by the affinity of the respective Fermi surface (FS) topologies. A selective match for one spin sign of the exchange-split FS in the ferromagnet with the FS of the paramagnet hints at the desired spin-polarized transmission. The reflectivity of an interface can suitably be detected via quantum well (QW) states. Demonstrations of their feasibility have thus far been limited to non-magnetic films, as observed in Cu [10,11], Ag [10,12,13] and Al [14]. For the ferromagnetic thin films of interest, spin-selective reflection of spin-up (*majority*) and spin-down (*minority*) electrons at the nonmagnetic substrate interface depends on how the respective wave functions are continued. In a nearly ideal spin filter, this will lead to QW states of only one spin type.

In this Letter, we report on the generation of spin-polarized standing waves by specific FS matching. The choice of a ferromagnetic Fe film on a W(110) substrate is guided by density functional theory (DFT) calculations. The FS is mapped by means of angle-resolved photoemission (ARPES). Fully spin-polarized QW states at $E_F$ are observed, yet exclusively in the *majority* FS sheets while minority states are not reflected. The data demonstrate the modification of Fermi level spin states in thin films, and the strong spin asymmetry documents the importance of a matched interface as a spin mirror.

The relevant features of the FS of bcc Fe as well as W have been derived with a modern DFT algorithm. The calculation employing the computer code WIEN2k [15] includes scalar-relativistic effects and the generalized gradient approximation (GGA) [16] to describe exchange and correlation. For the propagation of electrons perpendicular to the (110) interface, the (110) FS projections of the two materials are of importance.

The corresponding results are shown in Fig. 1(a) for W and in Fig. 1(b) for Fe. The W FS consists of an electron octahedron at the Γ point and hole octahedra at the six H points of the Brillouin zone (BZ). Fe carries two more d-electrons compared to W which leads to additional partially filled bands. The striking fact is that the *minority* FS of Fe in Fig. 1(b) has a high resemblance to that of W. In contrast, the *majority* FS exhibits as key features a much larger electron octahedron at Γ and tubular hole FS sheets connecting the H points. Both of these features find no match in the W FS. One must therefore conclude that, at least close to $E_F$, only *minority* states find an extension in the W substrate, whereas most of the *majority* electrons are confined to the ferromagnetic layer and cannot propagate across the interface.

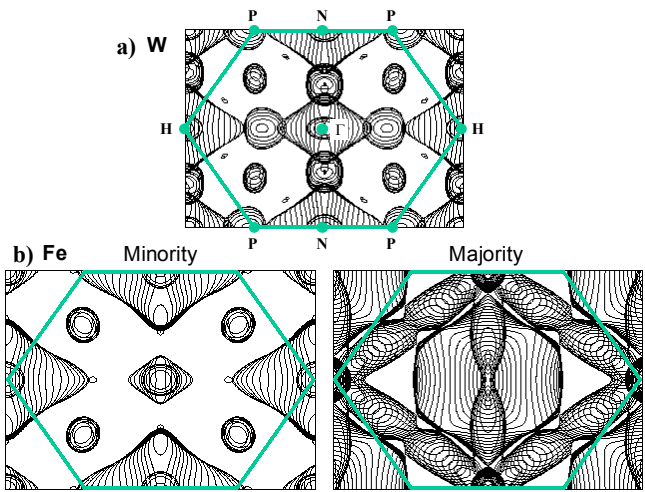

FIG. 1. Fermi surface (110) projections, calculated with DFT. (a) Fermi surface of W, with symmetry labels of basal plane. (b) Minority and majority FS of ferromagnetic iron (Fe BZ is ~9% larger than that of W). The *minority* FS consists of an electron pocket at Γ and hole pockets at H, its topology being similar to tungsten. The *majority* FS with its tubular hole sheets is rather dissimilar to W.

Experimentally, bcc Fe(110) films of high purity were grown by electron-beam evaporation onto a W(110) substrate and annealed at 500° C. The thickness range determined with W 4f core level attenuation was typically 20 – 40 Å, equivalent to 10 – 20 ML. The lattice constant of Fe is smaller than W by 9.2 %, yet structural defects rarely occur [17]. Instead, the strain is accommodated in approximately the first five layers by periodic elongations of the Fe lattice positions. Thus the electron density is not perturbed much at the interface. ARPES was performed at T = 25 K at beamline 7.0.1 of the Advanced Light Source in Berkeley with a momentum resolution of ~0.012 Å$^{-1}$ and a total energy resolution of ~35 meV. FS cross sections are determined for given photon energy (i.e. fixed $k_\perp$) by scanning the angle space in $k_x$ and $k_y$ direction. With the knowledge from the calculated FS, spin character can be assigned to the sheets without the necessity of a spin-resolved detection.

The experimental FS cross section for a ~ 15 ML film is displayed in Fig. 2, showing a section through the BZ close to the Γ point (hν = 128 eV). Both minority and majority FS sheet at Γ (small diamond and large hexagon, respectively) are intersected. Measured at T = 25 K far below the Curie temperature of $T_C$ = 1043 K, the FS is fully exchange-split. The calculation of Fig. 1(b) re-flects the maximum circumference of the FS sheets assumed at Γ, in excellent agreement with the ARPES data. An additional observation is the splitting of the *majority* contour into two sub-contours (see arrows), corresponding to QW states.

Quantization of the electron states results from the formation of ~ 10-20 discrete $k_z$-values perpendicular to the surface, corresponding to the limited number of monolayers. The wave vector of a standing wave with n nodes in a film of thickness d is given by $k_n = (\pi/d)\cdot n$. This neglects an additional phase factor that can result from the behavior of the wave function at the QW boundary [12]. Regarding states at $E_F$, the dispersion in the two other spatial directions remains intact. Each quantum number n selects a plane in the BZ parallel to the interface which slices through this FS topology.

The simultaneous observation of QW contours results from momentum broadening. For ARPES at given perpendicular momentum $k_\perp$ the limited electron escape depth λ leads to a broadening of $\Delta k_\perp \sim 2\pi/\lambda$. The electronic $k_n$-values themselves are also slightly broadened owing to the uncertainty relation by $\Delta k_n = 2\pi/d$. However, the escape depth λ ~ 5 ML is smaller than the film thickness of e.g. d ~ 15 ML. Therefore the ARPES depth sampling window is the dominant effect. Accordingly we expect to see 2-3 FS contours simultaneously, as is observed, with the specific observation depending on the curvature of the FS sheet.

Other essential parts of the BZ have also been explored for QW formation. In Fig. 3(a), a FS cross section for d ~ 15 ML away from Γ and closer to the BZ boundary plane (hν = 112 eV) is shown. It intersects the nu merous tubular *majority* FS sheets that connect the H points (labeled "A"). A large rectangle around H is a *minority* sheet (labeled "B"). Only the *majority* sheets exhibit formation of QW states, indicated by arrows in Fig. 3(a). A typical band map relating to a line cut through "A" is plotted in Fig. 3(b). It shows multiples of the hole pocket bands approaching $E_F$. The rather parallel dispersion behavior is characteristic for QW states.

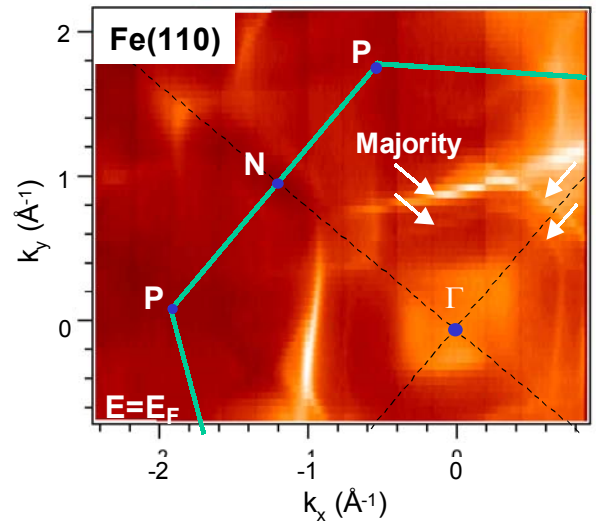

FIG. 2. Experimental ARPES Fermi surface of a Fe(110) film approx. 15 ML thick. The cross section with hν = 128 eV intersects the BZ close to the Γ point with *minority* (small) and *majority* (large) FS sheets. The majority FS sheet is split due to QW state formation.

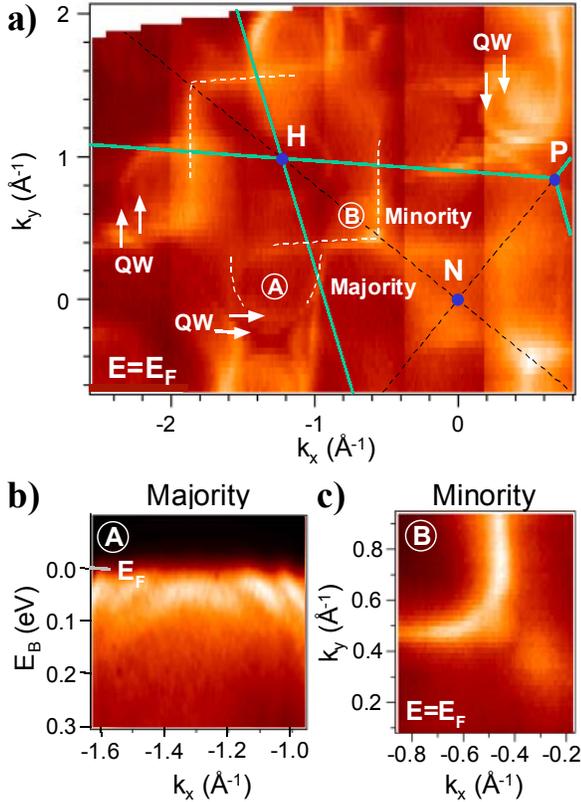

FIG. 3. (a) ARPES FS of Fe film ~ 15 ML thick with hν = 112 eV, intersecting the BZ close to the boundary plane. The numerous tubular *majority* sheets are intersected as circles (A). They exhibit QW multiplets. The *minority* sheet at H is intersected as square (B). (b) Bandmap (hν = 97eV) through majority sheet A showing QW band multiplet. (c) Minority sheet B at BZ boundary (hν = 97eV) measured with high statistics is devoid of QW states.

To ensure that minority QW states have not been overlooked, we varied photon energy and film thickness, yet with negative result. In Fig. 3(c) a close-up data set for a representative film of ~ 15 ML shows the minority FS around the H point, recorded exactly at BZ boundary plane (hν = 97 eV). Any QW states inside the hole pocket are clearly absent, within a detection limit of at least 1% peak intensity. Therefore, our experimental finding is that QW state formation at $E_F$ is exclusively observed for *majority* states. It proofs that the interfacial wave function match is drastically different for the two spin species. Intense QW states notably require that the interface causes hardly any loss of amplitude, thus smoothness on the atomic scale must be concluded.

The picture of high spin-polarization receives confirmation from a DFT calculation of the layered structure. It is modeled by 9 ML Fe on 9 ML W, numerically treated as indefinitely repeated in order to obtain a band structure. The lattice mismatch was circumvented by using the Fe lattice constant throughout, which best corresponds to the ARPES experiment where the relaxed Fe top layers are probed. In Fig. 4(a) a section of the multilayer structure is shown together with the magnetization density. Already near the interface Fe assumes a value of μ ~ 2.3 μ$_B$, virtually coinciding with the bulk value of 2.2 μ$_B$. W away from the interface cannot support a spin-splitting, yet at the interface W and Fe wave functions can intermix. A weak magnetization of *opposite* spin of the order of -0.1 μ$_B$ in the first W layer is consistent with a better penetration of minority states into W.

In Fig. 4(b) and 4(c) a representative spin-resolved band structure perpendicular to the interface is shown for a line containing the H point of the Fe BZ. The panels also indicate the relative contribution from Fe and W atomic orbitals, respectively. Key differences are observed between up and down spin states: the *majority* band structure in Fig. 4(b) clearly exhibits QW states, characterized by discrete and dispersionless energy levels. Concerning the population of these states, they are only formed by Fe valence electrons, while W electrons do not contribute perceptibly. Exactly at $E_F$ no state exists, unlike for bulk Fe, reflecting an altered FS density of states.

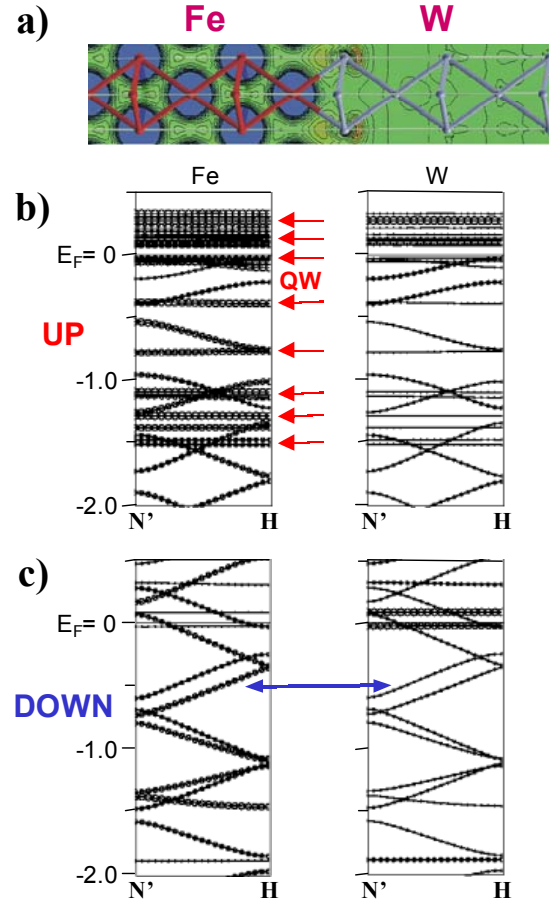

FIG. 4. DFT calculation of Fe-W system with 9 ML of Fe and W, respectively. (a) Magnetization density (blue), showing Fe magnetized up to the interface, while the adjacent W layer exhibits weak magnetization reversal (red). (b) Band structure perpendicular to the interface, with contribution of Fe or W wavefunction, respectively, indicated by circle size. *Majority* states form dispersionless QW states and are exclusively populated by Fe electrons. (c) Minority states disperse freely, with contribution from both Fe and W wavefunctions.

In contrast, the *minority* electrons in Fig. 4(c) are characterized by a dispersing band structure. Two bands known from bulk Fe are seen backfolded due to the supercell. Still they show a steep dispersion, and both bands cross the Fermi level. The bands are formed by both Fe and W orbitals, with the W valence electron density being less because of only 4 rather than 6 available d-electrons. The band calculation thus exemplifies the exclusive formation of *majority* QW states, consistent with the ARPES data. It also illustrates that the spin polarization, at least for selected parts of the BZ, can be virtually complete.

In comparing Fe/W to other well-studied interfaces, the most closely related system is Fe/Cr. The FS of bcc Cr is qualitatively similar to W. However, its larger BZ dimensions offer increased overlap with Fe *majority* states. A second effect is the antiferromagnetic ordering in Cr. Concomitantly it forms a spin density wave, the transition in bulk Cr occurring slightly above room temperature. As a result, large sections of the FS are gapped [18], thereby no longer providing a continuation for Fe minority wave functions. This conjecture would suggest a lesser degree of spin filtering compared to Fe/W, though this awaits experimental exploration.

Considering other elemental ferromagnets that can form part of a spin valve, fcc Co and Ni are usually combined with Cu interlayers. The FS of Cu resembles the *majority* FS of both these ferromagnets. However, their *minority* FS is not much different in extension, and therefore the selective advantage for spin transmission cannot be large. Accordingly, QW states of *both* signs have been detected in Co on Cu(100) by inverse photoemission [19]. This argument is refined by accurate calculations [4,6], predicting a GMR for defect-free layers of at most 90 % for Co/Cu multilayers, compared to 230–700 % for Fe/Cr, in support of the FS viewpoint.

The fully spin-polarized FS standing waves observed in ARPES are direct proof of the concept of selective matching of Fermi surfaces for one spin sign. The experimental results also underline the essential role of interfacial transmission for spin transport. It follows that it should be possible to optimize the spin-polarized yield in transport applications by tailoring the FS via doping and alloying. The FS data with quantized contours are furthermore direct evidence that the spin density of states at $E_F$ is considerably modified in these thin films. The ARPES data imply that the majority density of states at $E_F$ oscillates with thickness, and magnetic properties should thus be modulated. Such oscillations are indeed found for Fe films in the monolayer regime for the magnetic moments [20] and derived properties such as $T_C$ [21]. It is a fascinating outlook that bcc Fe can be grown on GaAs [22,23], and a Fe-based spin-mirror might be of interest for spin injection into the semiconductor.

The authors are grateful to A. Bostwick and H. Koh for technical support. This work was supported by the DFG (grant CL 124/3-2 and SFB 484), and by the BaCaTeC program.